\documentclass[sigconf]{acmart}
\usepackage{graphicx} 
\usepackage{subcaption}
\usepackage{wrapfig}

\setcopyright{none} 
\renewcommand\footnotetextcopyrightpermission[1]{}  

\acmConference[Augmented Educators and AI(CHI 2025 Workshop)]{Augmented Educators and AI: Shaping the Future of Human-AI Collaboration in Learning on CHI 2025 Workshop}{April 26,2025}{Yokohama, JAPAN}

\begin{document}

\title[Facilitating Instructors-LLM Collaboration for Problem Design]{Facilitating Instructors-LLM Collaboration for Problem Design in Introductory Programming Classrooms}

\author{Muntasir Hoq}
\affiliation{%
  \institution{NC State University}
  \city{Raleigh}
  \state{NC}
  \country{USA}}
\email{mhoq@ncsu.edu}

\author{Jessica Vandenberg}
\affiliation{%
  \institution{NC State University}
  \city{Raleigh}
  \state{NC}
  \country{USA}}
\email{jvanden2@ncsu.edu}

\author{Shuyin Jiao}
\affiliation{%
  \institution{NC State University}
  \city{Raleigh}
  \state{NC}
  \country{USA}}
\email{sjiao2@ncsu.edu}

\author{Seung Lee}
\affiliation{%
  \institution{NC State University}
  \city{Raleigh}
  \state{NC}
  \country{USA}}
\email{sylee@ncsu.edu}

\author{Bradford Mott}
\affiliation{%
  \institution{NC State University}
  \city{Raleigh}
  \state{NC}
  \country{USA}}
\email{bwmott@ncsu.edu}

\author{Narges Norouzi}
\affiliation{%
  \institution{University of California, Berkeley}
  \city{Berkeley}
  \state{CA}
  \country{USA}}
\email{norouzi@berkeley.edu}

\author{James Lester}
\affiliation{%
  \institution{NC State University}
  \city{Raleigh}
  \state{NC}
  \country{USA}}
\email{lester@ncsu.edu}

\author{Bita Akram}
\affiliation{%
  \institution{NC State University}
  \city{Raleigh}
  \state{NC}
  \country{USA}}
\email{bakram@ncsu.edu}

\renewcommand{\shortauthors}{Hoq et al.}

\begin{abstract}
Advancements in Large Language Models (LLMs), such as ChatGPT, offer significant
opportunities to enhance instructional support in introductory programming courses. While
extensive research has explored the effectiveness of LLMs in supporting student learning,
limited studies have examined how these models can assist instructors in designing
instructional activities. This work investigates how instructors' expertise in effective activity
design can be integrated with LLMs' ability to generate novel and targeted programming
problems, facilitating more effective activity creation for programming classrooms. To achieve
this, we employ a participatory design approach to develop an instructor-authoring tool that
incorporates LLM support, fostering collaboration between instructors and AI in generating
programming exercises. This tool also allows instructors to specify common student mistakes
and misconceptions, which informs the adaptive feedback generation process. We conduct case
studies with three instructors, analyzing how they use our system to design programming
problems for their introductory courses. Through these case studies, we assess instructors'
perceptions of the usefulness and limitations of LLMs in authoring problem statements for
instructional purposes. Additionally, we compare the efficiency, quality, effectiveness, and
coverage of designed activities when instructors create problems with and without structured
LLM prompting guidelines. Our findings provide insights into the potential of LLMs in enhancing instructor workflows and improving programming education and provide guidelines for
designing effective AI-assisted problem-authoring interfaces.
\end{abstract}



\maketitle

{\small
\noindent ©  This paper was adapted for the \textit{CHI 2025 Workshop on Augmented Educators and AI: Shaping the Future of Human and AI Cooperation in Learning},
held in Yokohama, Japan on April 26, 2025. This work is licensed under the Creative Commons Attribution 4.0 International License (CC BY 4.0).
}

\section{Introduction}
Advancements in Large Language Models (LLMs), such as ChatGPT, have resulted in quick access to classroom assistance, capable of streamlining instructors' preparation for classroom instruction. Recent work in the field suggests ways of integrating LLMs into introductory programming courses for the benefit of students, including problem design~\cite{denny2023trustaigeneratededucationalcontent}, feedback generation~\cite{phung2024automating} and providing real-time code explanations~\cite{sarsa2022automatic}. In this work, we investigate how instructors and LLMs collaborate to facilitate problem-design processes to promote active learning in introductory programming classrooms. Our goal is to facilitate the integration of instructors' expertise and intuition in designing practice problems for introductory programming classrooms with LLMs' ability to generate numerous programming activities within different contexts. We situate our experiments within INSIGHT AI-driven classroom assistant that features an instructor authoring tool enabling instructors to develop programming questions and share them with students and a student app that provides students with personalized feedback while working with those problems.
First, we explain the design and development of the instructor-authoring tool, which enables seamless collaboration between instructors and LLMs to design introductory programming questions. As part of the authoring process instructors work with LLMs to propose possible novice-generated solutions and highlight potential underlying misconceptions in those solutions. The instructors can further co-author relevant feedback with LLMs to be propagated to students who demonstrate evidence of those misconceptions in their code. We then show case studies of instructors utilizing these tools to generate programming activities for their classrooms. 

We have instructors design activities with and without guidance from the research team. The guidance involves providing instructors with exemplary prompts and general guidelines on effective ways to interact with LLMs through carefully crafted prompts for generating problem statements, sample solutions with misconceptions, and feedback. We then obtained instructors' opinions on the effectiveness, efficiency, and coverage of the learning objectives of the activity design process and the results.

To understand the current pedagogical needs and challenges faced by introductory programming instructors, we pose and explore the following research questions:
\begin{itemize}
    \item RQ1: How do instructors perceive the usefulness and limitations of LLMs in authoring problem statements, generating possible solutions, and providing feedback for instructional purposes?
    \item RQ2: How do instructors evaluate the effectiveness and efficiency of LLM-generated outputs (problem statements, solutions, and feedback) in unguided versus guided approaches?
    \item RQ3: To what extent do instructors trust and feel confident in integrating LLM-generated problem statements, solutions, and feedback into their instructional practices?
\end{itemize}

Our results demonstrate the effectiveness of the guidance in improving the effectiveness and efficiency of the generated activities, while some instructors expressed concern about the reduction in creativity levels when following the guided approach. 


\section{Related Work}
\subsection{LLMs for Instruction}
LLMs have recently gained prominence as powerful tools in education, particularly for programming instruction ~\cite{prather2023robots}. Denny et al. ~\cite{denny2023trustaigeneratededucationalcontent, denny2024computing} examine how AI-driven technologies enable the creation of diverse programming exercises tailored to individual learners' needs. These automatically generated resources help educators enhance their teaching while providing students with personalized practice. LLMs have also been used to generate various types of instructional materials, such as programming exercises ~\cite{denny2022robosourcingeducationalresources, 10.1145/3501385.3543957, ta2023exgen}, worked examples ~\cite{jury2024evaluating}, programming multiple-choice questions ~\cite{doughty2024comparative, tran2023generating}, and learning objectives ~\cite{sridhar2023harnessing}. Findings suggest that LLMs can significantly reduce the time instructors spend creating these resources.
\subsection{Instructor Authoring Tools} 

LLM-based approaches have been explored to assist instructors in creating teaching materials ~\cite{raihan2025large}. Some methods leverage tools like ChatGPT directly, using various prompting strategies without accessing OpenAI’s APIs ~\cite{sarsa2022automatic, del2024evaluating}. Others develop custom authoring tools that provide instructor-facing interfaces while integrating LLM APIs on the backend. Jordan et al. ~\cite{matelsky2023largelanguagemodelassistededucation} introduce FreeText, an LLM-powered tool that helps educators design and assess open-ended questions. Instructors can generate questions via an HTTP API or a web-based interface, with optional assessment criteria to iteratively refine prompts. Across these different approaches, LLM-based tools have been found to support educators in crafting clearer, more effective questions with less effort. However, they do not replace human instructors, as they lack pedagogical judgment, may introduce biases in AI-generated refinements, and still require instructor oversight to ensure accuracy and alignment with learning objectives.


\section{INSIGHT Classroom Assistant}

INSIGHT classroom assistant is an AI-driven tool that features an instructor authoring tool and a student-facing app.
This tool enables instructors to design and share programming promotes with students while providing students with adaptive feedback. 
\subsection{INSIGHT Authoring Tool}
\begin{figure*}[t]  
    \centering
    \begin{minipage}{0.5\textwidth}  
        \centering
        \begin{subfigure}{\textwidth}
            \centering
            \includegraphics[width=\textwidth]{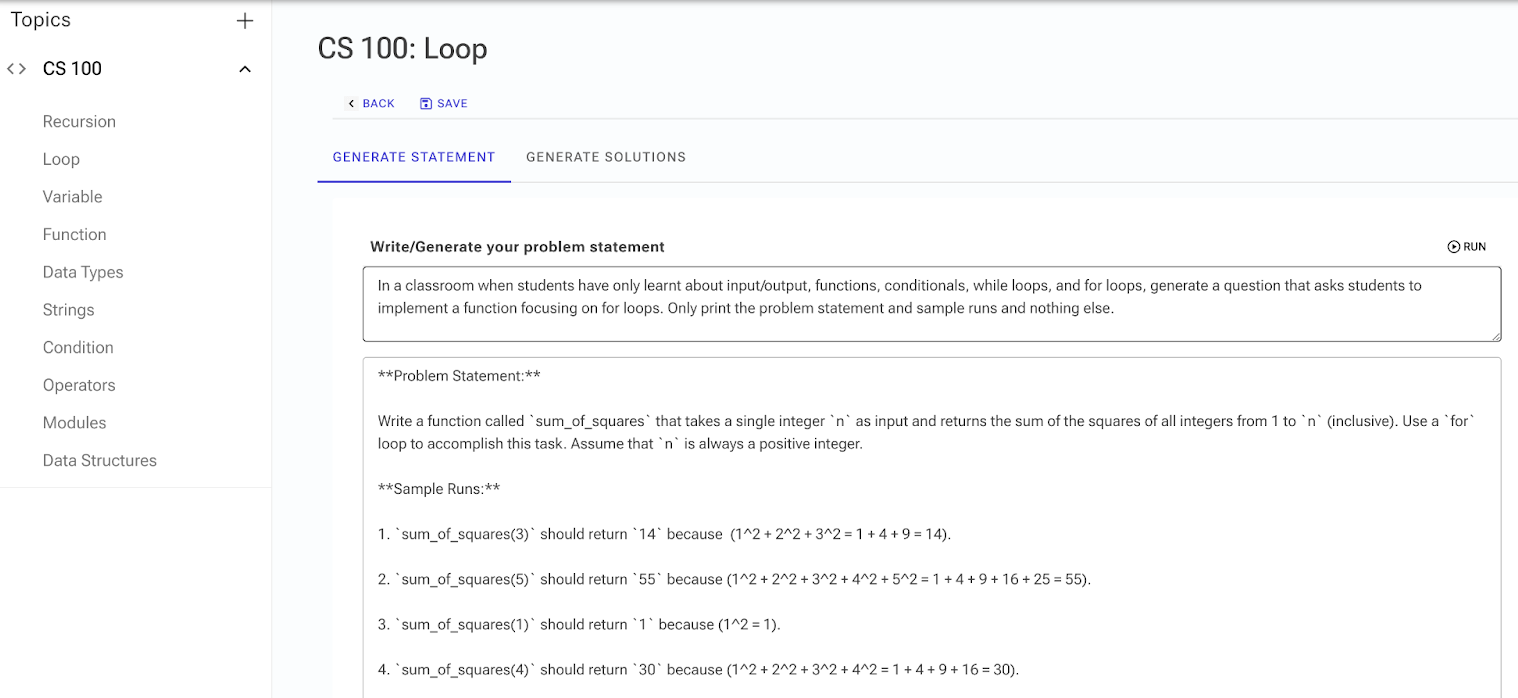}
            \subcaption{Problem generation.}
            \label{subfig:a}
        \end{subfigure}
        \begin{subfigure}{\textwidth}
            \centering
            \includegraphics[width=\textwidth]{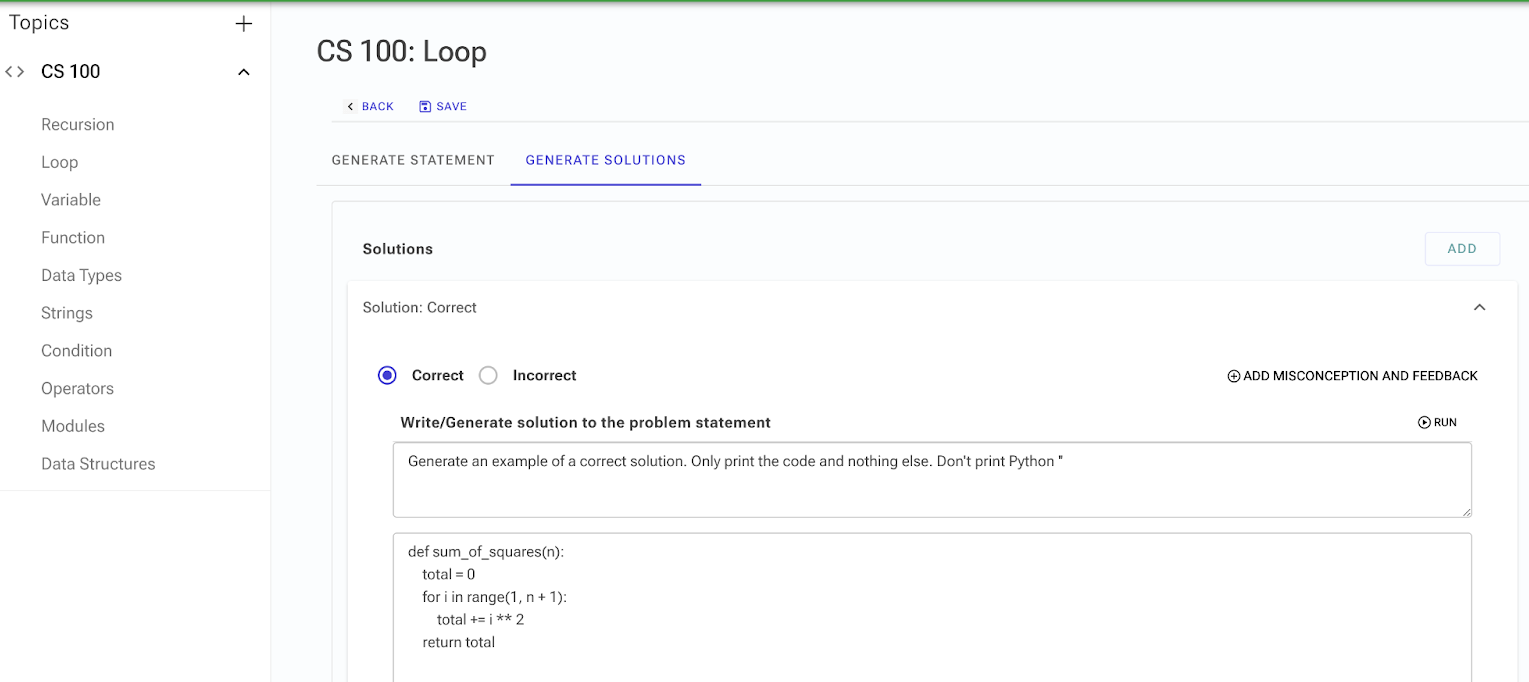}
            \subcaption{An incorrect solution generation.}
            \label{subfig:b}
        \end{subfigure}
    \end{minipage}
    \hfill
    \begin{minipage}{0.48\textwidth}  
        \centering
        \begin{subfigure}{\textwidth}
            \centering
            \includegraphics[width=\textwidth]{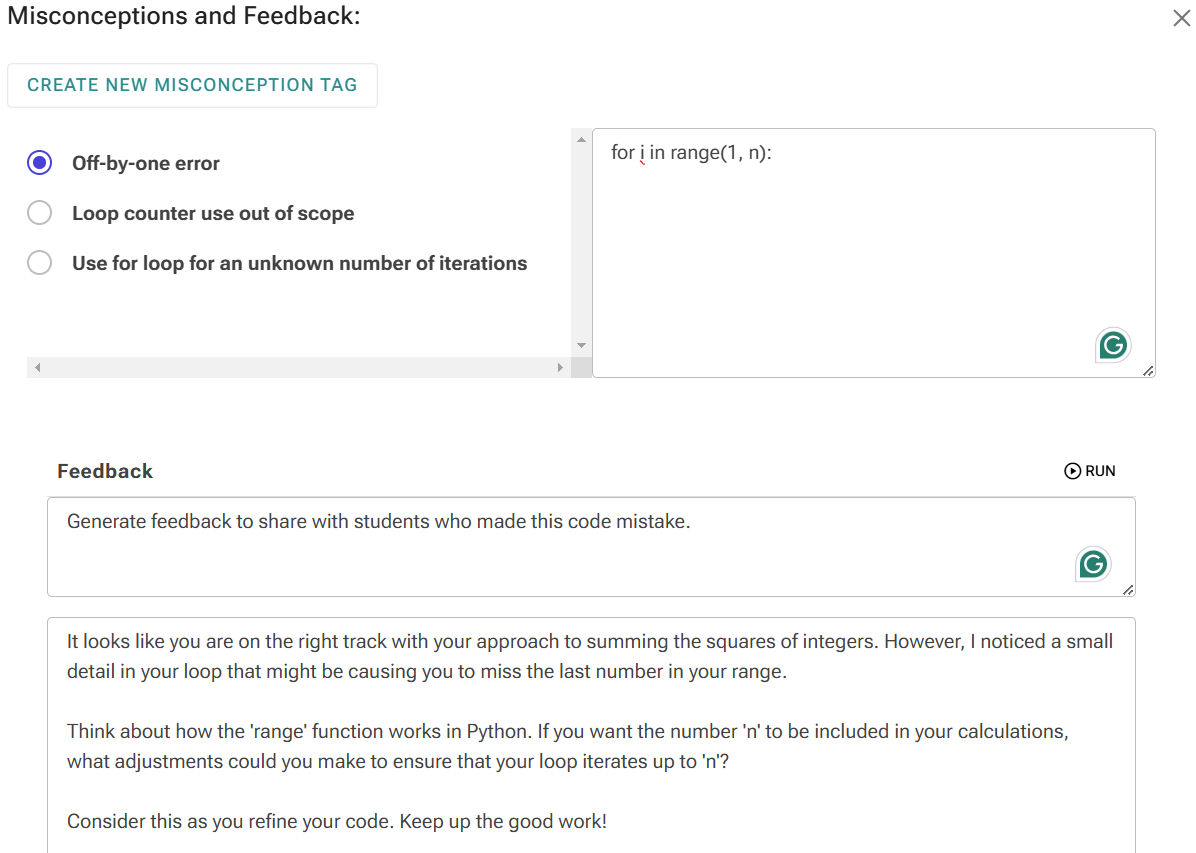}
            \subcaption{Misconception selection and feedback generation.}
            \label{subfig:c}
        \end{subfigure}
        \begin{subfigure}{\textwidth}
            \centering
            \includegraphics[width=\textwidth]{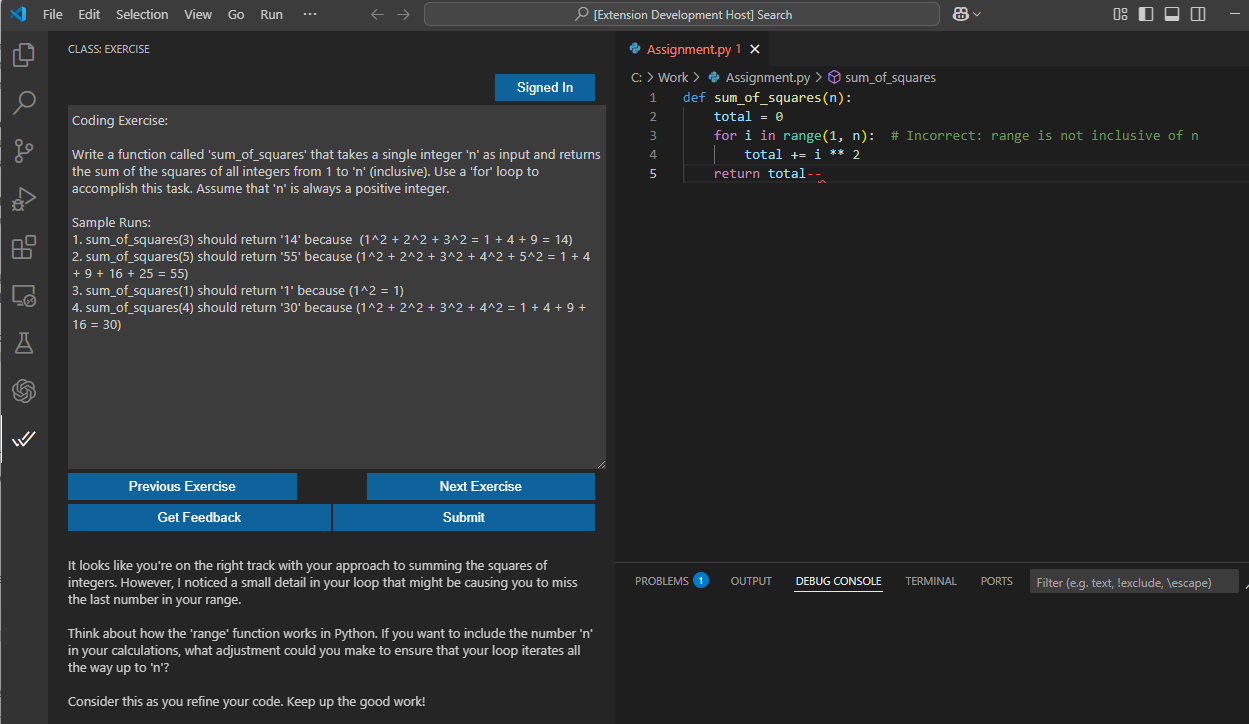}
            \subcaption{Student app as a VS Code extension.}
            \label{subfig:d}
        \end{subfigure}
    \end{minipage}

    \caption{Different scenarios in the authoring tool interface and the student app interface in VS Code.}
    \label{fig:four_scenarios}
\end{figure*}

The INSIGHT authoring tool enables instructors to create topic-tailored coding exercises that are specific to the course. All coding exercises are categorized by their course subject and topics in the tool, along with corresponding solutions and feedback. The authoring tool allows instructors to add new coding exercises and provide sample solutions with common misconceptions about the exercises. The tool also allows instructors to leverage Large Language Models (LLMs) in generating coding exercises, solutions, and feedback. In this way, instructors can incorporate their pedagogical expertise with the extensive problem and solution spaces of LLMs to generate the most relevant exercises for students.Figures~\ref{subfig:a},~\ref{subfig:b} and~\ref{subfig:c} demonstrate the problem design interface, solution generation interface, and misconception selection and feedback generation interface.

\subsection{Student-Facing App}
To allow students to receive the instructor-sent coding exercises and write their own solutions within the interface, we have developed the INSIGHT student-facing app. The student app enables students to access activities within the app, such as viewing coding exercises sent in the class, receiving immediate personalized feedback and hints to develop solution code, and submitting their solutions in real time~\cite{hoq2024towards,niousha2024use}. The initial version of the INSIGHT student app is developed as a Visual Studio Code extension as illustrated in Figure~\ref{subfig:d}. 

\section{Methodology}
\subsection{Recruitment and Participants}
We used a purposive sampling approach~\cite{kelly2010qualitative} to identify instructors at a large, public R-1 university in the United States who had experience teaching introductory programming (CS1) courses. Potential participants were approached through individual outreach, focusing on colleagues familiar with designing CS1 materials. Three instructors who met the criteria agreed to participate, each having taught CS1 for at least one semester and had been responsible for designing programming problems, solutions, and feedback in their courses. All participants provided informed consent, and their identities were anonymized to protect confidentiality. Although the sample size is small, it facilitated an in-depth qualitative exploration of their practices and perceptions when leveraging LLMs for CS1 instructional design~\cite{palinkas2015purposeful}.

\subsection{Protocol Development}
The protocol for this study was developed through an iterative process that incorporated feedback from subject-matter experts. Our aim was to create a structured yet flexible approach that would allow CS1 instructors to collaborate with an LLM—--specifically ChatGPT--—through our authoring tool to design programming problems, solutions, and feedback that target key learning objectives.

The central objectives and research questions of the study were crafted based on the evolving role of LLMs in educational settings, focusing on instructor-LLM collaboration in CS1 problem generation. The \textbf{three main steps} of the study were as follows:

\begin{itemize}
    \item Problem Statement Generation – How instructors conceptualize and create new CS1 programming problems and how LLMs can facilitate that process.
    \item Solution Formulation – How instructors generate, refine, and evaluate a range of both correct and incorrect solutions collaboratively with LLMs.
    \item Feedback and Misconception Tagging – How instructors identify common misconceptions in incorrect solutions and produce constructive feedback with the help of LLMs.
\end{itemize}

\subsubsection*{\textbf{Designing a Two-Round Procedure: Unguided vs. Guided: }}

To explore how a guided and instructed approach can help instructors use LLM and improve the authoring experience in terms of quality and efficiency over an unguided and independent approach, we designed two separate rounds for each instructor.

\begin{figure}  
    \centering
    \includegraphics[width=\columnwidth]{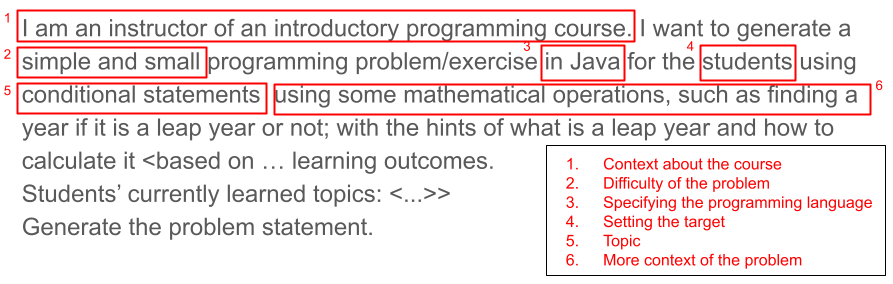}
    \caption{Example of guided problem generation.}
    \label{fig:guided}
\end{figure}

\begin{itemize}
    \item Unguided Round – Instructors were asked to interact with ChatGPT without any structured prompts or explicit guidance. The goal was to observe and document instructors’ natural strategies when using an LLM to generate a new programming problem statement, produce multiple solutions (correct and incorrect), identify misconceptions in the incorrect solutions, and propose feedback.
    \item Guided Round – Instructors were provided with example prompts offering a clear structure and recommendations for prompt engineering. They were then asked to repeat the three main tasks (problem generation, solutions, and feedback) using the more guided approach. The prompt examples and instructions were curated through pilot testing and expert feedback. The purpose of this ``guided'' round was to help the instructors collaborate with LLMs more effectively and see how the guidance can improve the quality, efficiency, and effectiveness of the authoring process. Instructors used those structured prompts and modified them if needed in this round. The guided instructions included providing accurate and effective context to LLMs in the problem, solutions, and feedback generation process. The exemplary prompts provided a structured framework that specifies key elements such as concepts, learning outcomes, difficulty level, target student audience, and other contextual factors to guide the generation of problem statements. Similarly, guidance for the solution and feedback generation process included getting good coverage of student mistakes while generating incorrect solutions, specifying the structure and level of detail of the generated feedback without compromising precision or enabling expedient help-seeking~\cite{hoq2024detecting}, and tagging useful misconception names. Figure~\ref{fig:guided} illustrates one example prompt for the problem generation steps.
\end{itemize}

By comparing these two rounds, the protocol aimed to uncover differences in instructors’ perceptions of effectiveness and ease of use between the ``unguided'' and ``guided'' approaches.

\subsection{Interview Procedure}
Instructors were first given an overview of the INSIGHT Classroom Assistant and its goals.  They were then instructed to author two questions, possible correct and incorrect solutions, and misconceptions and feedback assigned to each incorrect solution. In the first design round (unguided), instructors were to interact openly with LLMs to design all the required materials. In the second design round (guided), they received instructions and examples on the efficient usage of LLMs for problem, solution, and feedback generation. 

Before starting the first (``unguided'') round, instructors were asked questions about their current strategies in designing programming problems, generating possible solutions and feedback for incorrect solutions (see Section \ref{subsection:traditional}). Following completion of both rounds (unguided and guided), instructors engage in a final interview to understand differences in effectiveness, efficiency, and coverage of learning objectives between the unguided and guided rounds (see Section \ref{subsection:unguided_vs_guided}). Instructors also expressed their overall satisfaction and general perceptions with LLM assistance in problem, solution, and feedback generation (see Section \ref{subsection:general_perception}).

\subsection{Ethical Considerations and Data Management}
University ethics review board approval was sought as part of conducting the research project, which included in-depth instructor interviews. Instructors provided informed consent via email before the interviews and verbally reaffirmed their consent at the start of each session. Each interview was audio and video recorded, and the recordings were transcribed automatically using speech-to-text software. Subsequently, an author reviewed and corrected these transcripts to ensure accuracy and completeness. All data, including records, transcripts and related artifacts, were stored securely in a password-protected environment, accessible only to authorized research personnel.

\subsection{Analysis}
Qualitative analysis of interview transcripts followed a rigorous thematic analysis approach to identify and organize key patterns within the data \cite{braun2012thematic}. The process began with an open coding phase, where responses were systematically reviewed, and initial codes were generated to capture recurring ideas and variations in participants' responses \cite{vaismoradi2013content}. These codes were iteratively refined and synthesized into broader themes that encapsulated meaningful patterns within the data. Throughout the analysis, the research team engaged in structured analytic discussions to critically examine and refine emerging themes, ensuring they were well-supported by the data and aligned with the research objectives. This systematic and reflexive approach facilitated a nuanced interpretation of participant perspectives and maintained coherence across the identified themes~\cite{clarke2017thematic}.

\section{Results}
This section presents the key themes that emerged from instructor interviews regarding their experiences using LLM for problem statement generation, feedback, and misconceptions. The findings are organized into two sections: (1) instructors' general perceptions of using LLM in the classroom, including their trust and confidence in its use, and (2) instructors' reflections on the unguided and guided approaches they experienced during the interview. The unguided approach refers to instructors using ChatGPT independently, as they normally would, while the guided approach refers to their use of ChatGPT with scaffolded prompting provided by the research team.

\subsection{Instructors' Traditional Methods for Designing Problems}
\label{subsection:traditional}
Instructors discussed how they traditionally create problem statements, feedback, and misconceptions, largely relying on their own expertise and past experiences. This approach ensured full control over problem design but required additional time to maintain coverage and variety. One instructor noted, “I try to just develop those programming questions myself, based on my experience,” while another emphasized structuring problems around specific learning outcomes. A key advantage of this approach was the ability to create engaging, context-rich problems tailored to student experiences. One instructor highlighted how TAs contributed problem ideas relevant to student life, while another stressed the importance of incorporating narratives to provide motivation and real-world context.

\subsection{Instructors’ General Perceptions of Using AI in the Classroom}
\label{subsection:general_perception}
Instructors saw LLMs as a useful tool for \textbf{brainstorming} and expediting the problem-generation process but remained skeptical of its accuracy and appropriateness for direct student use. One instructor noted, “It’s really helpful... I do not have to start from scratch. I can just use or be inspired by LLM’s ideas.” However, despite these advantages, instructors emphasized the need for human oversight. Another instructor pointed out, “You cannot guarantee it's correct, so you have to be very careful to make sure it's 100\% accurate.”

Despite these concerns, some instructors—--particularly those newer to teaching--—found LLMs useful as a \textbf{starting point for problem development}. One instructor mentioned that for those with less experience, LLM could be a \textbf{time-saving and creativity-enhancing tool}, noting, “For a very entry-level instructor with little or no experience, the [guided] approach is definitely helpful.” Another added that while LLM-generated questions lacked depth, they were still a \textbf{useful foundation for refinement}: “The questions don’t have a whole lot of soul to them… but they’re useful.” Another instructor described the issue of over-reliance on generic problem templates, saying, “If the LLM produces too ‘classic’ or widely published problems, students may find them with a quick search, reducing the challenge and uniqueness.”

When it came to trusting LLM-generated solutions, instructors were generally cautious. One instructor stated that LLM-generated content needed thorough verification, explaining, “The code generated by ChatGPT may be incorrect or untested; I feel a strong need to verify the correctness by running or unit-testing the solutions.” 

Instructors also mentioned several challenges for using LLMs as a direct tool to provide students with feedback. One of the biggest concerns was \textbf{over-detailing} in ChatGPT-generated feedback. Multiple instructors noted that ChatGPT often provided too much explicit guidance or even full solutions instead of strategic hints. One stated, “A lot of the time, it will give way too much explicit guidance on how to improve your code, which is not what I’m aiming for.” Another echoed this sentiment, explaining that LLM-generated feedback must be filtered before reaching students: “I can never trust directly giving students anything. We need a middle layer—you cannot directly expose ChatGPT to students at this stage.”

\subsection{Comparing the Unguided and Guided Approaches}
\label{subsection:unguided_vs_guided}
When comparing their experiences using ChatGPT independently (unguided) versus using it with scaffolded prompts (guided), instructors generally found the guided approach to be more efficient and structured, though some still valued the flexibility of the unguided approach for brainstorming.

\subsubsection{Efficiency and Coverage}
The guided approach was notably faster for generating structured problem statements and feedback. One instructor explained, “The guided way was probably faster. In my case, I’m not sure that would hold true for more experienced instructors.” Another added, “The second approach is definitely helpful... It helps me cover something I missed.” The scaffolded prompts provided clear direction, reducing the time spent refining AI-generated responses.

In addition, instructors found that the guided approach led to better coverage of learning objectives. One explained that providing structured prompts improved problem relevance: “Being able to provide a list of learning outcomes and have it generate questions as well as what learning outcomes they satisfy would be useful.” Another instructor described how the guided approach allowed for better alignment with learning goals without extensive back-and-forth revisions, stating, 'The guided approach results in problem statements and solutions that better match my needs without requiring so many refinements.

\subsubsection{Creativity and Engagement}
While the guided approach improved efficiency, some instructors felt that the unguided approach allowed for more creative problem formulation. One instructor found the unguided approach useful for brainstorming: “I was just like really brainstorming, throwing spaghetti at a wall.” Another pointed out that while the guided approach was more structured, the unguided approach sometimes led to novel problem ideas: “The problem statements produced by the guided way came out a bit more effective at meeting the exact requirements… but the unguided way produced more interesting problems that needed a lot of fleshing out.”

However, the effectiveness of the unguided approach depended on the instructor’s familiarity with ChatGPT. One instructor mentioned that experience using ChatGPT played a significant role in whether the unguided approach was effective, stating, “Instructors need to be familiar to be effective and already have some experience using ChatGPT to generate ideas and solutions.” Without prior experience, some found the unguided approach time-consuming, as one instructor explained: “It required multiple iterations of prompts to arrive at suitably tailored problems or feedback.”

\subsubsection{Effectiveness of Feedback Generation}
When generating feedback, the instructors found the guided approach more structured and reliable, although it still required manual oversight. One instructor described how ChatGPT, when prompted with specific guidelines, helped customize feedback to different possible student responses: “The second version (guided) helped me more to come up with different possible solutions and then customize feedback for those specific versions.” Another instructor found that the guided approach improved the specificity of ChatGPT-generated feedback but still had reservations about its quality: “I think the guided feedback was a lot better. But I think that’s also partly because I didn’t really experiment with trying to write a better feedback prompt in the unguided case.”

However, the concern that ChatGPT-generated feedback was too detailed or explicit remained prominent. One instructor remarked, “It’s not what I want to give to the student.” This remark was common in both the unguided and guided approaches, indicating the general limitations of LLM-generated feedback regarding reliability and faithfulness~\cite{jacobs2024evaluating,jia2024assessing,phung2023generative}. The real-time collaboration between instructors and LLMs for feedback generation can address this issue by bringing instructors into the loop. In the future, we will use mediatory layers using LLMs or explainable ML models~\cite{phung2024automating,hoq2023sann} to propagate instructor-verified feedback to students.


\section{Discussion}
Overall, instructors saw LLMs as a useful tool for problem generation, feedback, and misconceptions, but they remained cautious about fully trusting LLM-generated content without human oversight. While the guided approach provided efficiency, structure, and better alignment with learning objectives, some instructors still valued the unguided approach for its flexibility and potential to inspire more creative problem statements. The findings suggest that a hybrid approach, where instructors use structured LLM assistance while retaining control over refinement and customization, maybe the most effective way to integrate LLMs into the problem-authoring process.

While our study highlights the benefits and limitations of instructor-LLM collaboration for problem authoring, the small number of participants limits the generalizability of our findings. In the future, we aim to conduct a large-scale study to validate these insights across a variety of instructional settings (e.g., university type, classroom size, and teaching modalities). Furthermore, future research should explore how different levels of LLM guidance impact instructor efficiency and the quality of designed problems. Specifically, we aim to investigate the granularity of guidance by systematically varying how much control instructors have over problem generation. This includes comparing fully structured interactions (e.g., dropdown menus and predefined templates) with flexible, iterative LLM prompting to determine which approach best balances efficiency and creativity. Additionally, future work should examine how problem variation affects instructional effectiveness, including whether generating multiple versions of a problem at once leads to better learning outcomes compared to single-instance generation.  Another critical area is the role of LLMs in problem validation, where we will test whether automated feedback on instructor-authored problems improves quality compared to manual review alone. 

The results will be used to design an interface that facilitates instructor-LLM collaboration by balancing instructor flexibility and LLM involvement, ensuring that problem design is efficient and pedagogically effective.

\section{Conclusion}
In this study, we conducted case studies to explore how instructors can collaborate with LLMs to design programming problems for introductory programming classrooms in unguided and guided formats. Our findings highlight the potential of LLM-assisted authoring tools to facilitate problem design utilizing instructors' expertise and LLMs' generative capabilities. This work lays the foundation for a large-scale study that will systematically examine the factors influencing effective instructor-LLM collaboration. The insights gained will inform the development of an interface that enhances both the efficiency and pedagogical quality of programming problems.

\section*{Acknowledgments}
This research was supported by the National Science Foundation (NSF) under Grants DUE-2236195 and DUE-2331965. Any opinions, findings, and conclusions expressed in this material are those of the authors and do not necessarily reflect the views of the NSF.

\bibliographystyle{ACM-Reference-Format}
\bibliography{sample-base}

\end{document}